# Direct characterization of planar waveguide modes by Fourier plane fluorescence leakage radiation microscopy

Douguo Zhang[a], Qiang Fu, Xiangxian Wang, Pei Wang[b] and Hai Ming

*Institute of Photonics, Department of Optics and Optical Engineering, University of Science and Technology of China, Hefei, Anhui, 230026, P. R. China*

**Abstract**

In this letter, the leakage radiation microscopy (LRM) is extended into characterization of planar waveguide modes (WMs) rather than surface plasmon polaritons (SPPs) taking advantages of the coupling between WMs and fluorescence emission. Propagation constants of different WMs allowed in the same planar waveguide can be simultaneously and rapidly derived from the Fourier plane image of fluorescence based LRM. Numerical simulations are also carried out to calculate propagation constants of these modes, which are consistent with experimental results. Our experiments provide a simple but high efficient method to characterize planar waveguides.

Electronic mail: [a]dgzhang@ustc.edu.cn, [b]wangpei@ustc.edu.cn

Leakage radiation (LR) of SPPs is emitted from the interface between the metal film and dielectric medium of higher refractive index (ordinary glass substrate). The LR intensity at a certain lateral position on the glass side of the metal/glass interface is proportional to that of the SPPs on the metal side at the same lateral position. Further more, with respect to the normal of the interface; LR radiates at a characteristic surface plasmon resonance (SPR) angle. So excitation and propagation of SPPs can be characterized through the detection of LR[1, 2]. This is the working principle of LRM. Now LRM has been widely used in the field of plasmonics to characterize the wave-number and as well as propagation length or route of SPPs[3, 4]. Known to all, the mechanism of excitation of WMs and SPPs are the same, which is to realize the momentum matching between the incident light and the modes. WMs also can be excited with high refractive index prism, in reversely, LR of WMs will radiate back to prism in the same way as that of SPPs. In principally, optical properties of WMs also can be characterized by the LRM. In this letter, we demonstrate how to characterize the WMs with the LRM; both the wave-numbers and propagation lengths can be measured from the Fourier plane image of LRM. For convenience and accuracy, these WMs are excited by the Rhodamine B (RhB) molecules doped in the waveguide, which serve as the excitation source or gain medium [5, 6, 7]. To the best of our knowledge, there is no report on measure the propagation constants, especially the propagation lengths of WMs with the fluorescence based LRM. The advantages of the fluorescence based LRM over the common LRM will be discussed in the following experiment. Planar waveguide is one of the key elements in integrated optics, so characterization of WMs allowed in the waveguide is of highly importance.

The plane waveguide investigated in the experiment comprises of an RhB doped PMMA film spinning coated onto a silver film. The 45nm thick silver film is deposited on a glass substrate by electron beam deposition. Thickness of the PMMA layer is sensitive to modes supported by the four-layer film (Air- PMMA-Ag-Glass). The thin PMMA film only supports SPPs mode ($TM_0$), while the thick one allows more modes, such as $TM_1$, $TE_0$ and $TE_1$ and so on. In our experiment, thickness of the PMMA film is about 500nm. More details of the samples' preparation were described in Reference [8].

A linear polarized laser beam at 532nm wavelength was expanded by lens array and then stroked on the rear aperture of an oil immersed objective (60X, N.A. 1.42). Then it was tightly focused onto the PMMA film and excited the RhB molecules. The emitting fluorescence were collected by the same objective and directed to the CCD camera. A long pass edge filter (edge wavelength 532nm) was placed before the CCD camera to reject the excitation laser (532nm). Another band-pass filter was used to select the emission wavelength, whose center wavelength is chosen at 580nm. Full-width at half-maximum (FWHM) of the band pass region is 10nm. More details about the LRM were described in Reference [8, 9].

Figure 1 (a) shows the Fourier plane image of the reflected 532nm laser, which was captured without any filter in front of the CCD. Three pairs of symmetrical dark arcs appear in this image, which are typical signatures of excitation of WMs by linear polarized laser beam. Each pair of arcs in Figure 1 (a) is corresponding to one WM mode, which are $TM_1$, $TM_2$ and $TE_1$ respectively. The $TE_0$ mode does not appear due to the limited value of the N.A, which will be discussed in the following. The wave-numbers of the three modes are 1.36 $K_0$, 1.21 $K_0$, and 1.04 $K_0$ respectively estimated from the known N.A of objective[9].

In theory, propagation length of the WMs can be derived from the wavenumber distribution in Fourier plane, which is the intensity distribution along the diameter across these dark arcs. But, as shown in Figure 1 (a), only a small portion of incident light transfers to WMs and also many fringes arise in the image due to the coherence of laser beam. As a result, it will not be accurate to estimate the propagation lengths from the dark arcs.

Our previously experiments shows that excited fluorescence molecules can excite not only the SPPs, but also the WMs, which both result in the directional emission of fluorescence[9]. Excitation of WMs and SPPs by excited fluorescence molecules has the following advantages for the propagation constants' measurements: the first one, there is no need to do the optical alignment or fabricating nanostructures with high cost nanofabrication facility to meet the momentum matching. When the fluorescence molecules are excited, it can transfer to the WMs or SPPs. second, most of the emitting fluorescence transfers to the WMs or SPPs, so the background noise is very low, which makes it more accurate to dig out the information of these modes. The third, there are no coherence induced fringes. On the other hand, the broad spectrum of fluorescence is a disadvantage for the characterization because propagation constants are dependent on the excitation wavelengths. To solve this problem, a band pass filter is used to select the wanted wavelength for characterization.

Figure 1 (b) shows the Fourier plane image of emitting fluorescence at 580nm wavelength. There are four bright rings which is the typical waveguide mode-coupled emission (WMCE) and similar as the SPCE. The four bright rings are corresponding to four WMs existing in the plane waveguide. To qualitatively derive the propagation constants of these modes, pixels intensity distribution along the red-dash line across the bright rings is extracted as shown in Figure 2(a), which is also the wavenumber distribution of the excited WMs in Fourier plane. There are four pairs of peaks on this curve, and the peak to peak distance, for examples, d1=A2-A1, d2=B2-B1, and d3=C2-C1, represents the diameter of the corresponding ring. Based on the known N.A of the objective and the diameter of the bright rings, wavenumbers of these modes can be estimated as 1.42 $K_0$, 1.34 $K_0$, 1.169 $K_0$, 1.01 $K_0$ respectively, where $K_0$ is the wavenumber of the 580nm light in vacuum[7]. As far as the $TE_0$ mode is concerned, the effective index (1.42) is not less than N.A of the objective, so the LR of this mode can not be totally collected. As a result, the estimation of propagation constant of $TE_0$ mode will not be precise.

To derive the propagation length, left section of the curve in Figure 2 (a) with one peak (A1, B1, or C1) are extracted as shown in Figure 2 (b), (c), (d). These curves are wavenumber distributions of corresponding WM, and can be described with a Lorentzian type function from which FWHM encoding the losses or propagation length can be inferred[3, 10, and 11]. In Figure 2 (b), (c) and (d), the red dots are experimental data and the blue curves are Lorentzian curves. Propagation lengths of the three modes can be derived from FWHM of the Lorentzian curves, which are 7.936μm ($TM_1$), 14.51μm ($TE_1$), and 18.47μm ($TM_2$) respectively.

Numerical simulation is carried out to calculate the wavenumbers and propagation lengths of WMs existing in the four-layer film (Air-PMMA-Ag-Glass)[12]. The ATR curves are calculated based on Fresnel formula and boundary conditions, as shown in Figure 4. In the simulation, the incident light strikes onto the silver film from the glass substrate. Refractive index of silver is 0.121+3.579i (at 580nm), 0.129+3.193i (at 532nm)[13]. The refractive indices of the glass and PMMA are 1.52 and 1.50 respectively. Thickness of the PMMA and Ag film are 500nm and 45nm, which are the same as that used in the experiment. The valley positions of the ATR cures represent the wavenumber of the excited WMs. At the 532nm excitation wavelength, four WMs are excited and the wave numbers are 1.43 $K_0$, (TE$_0$ mode, out of the collection range of the objective, so it does not appear in Figure 1 (a)), 1.37$K_0$ (TM$_1$ mode), 1.22$K_0$ (TE$_1$ mode), and 1.06$K_0$ (TM$_2$ mode). At the 580nm excitation wavelength, the wavenumbers are 1.42 $K_0$ (TE$_0$ mode), 1.35 $K_0$ (TM$_1$ mode), 1.18 $K_0$ (TE$_1$ mode) and 1.01 $K_0$ (TM$_2$ mode). These values are consistent with experimental results derived from Fourier plane images.

The propagation constants $\beta_L$ of WMs can be described in Eq (1), which contains both the real and imaginary parts[14].

$$\beta_L = \beta_0 + \Delta\beta_L = [\text{Re}(\beta_0) + \text{Re}(\Delta\beta_L)] + i[\text{Im}(\beta_0) + \text{Im}(\Delta\beta_L)] \quad (1)$$

Where $\beta_0$ is the propagations constant of WM when the leakage radiation loss is not taking into account and only intrinsic loss is considered. $\Delta\beta_L$ is the item induced by leakage radiation loss. Both $\beta_0$ and $\Delta\beta_L$ are complex number. The real part of $\beta_L$ is direct ratio to the wavenumber and the imaginary part is an inverse ratio to propagation lengths of the WMs. In the region of resonance where WMs are excited, reflectivity can be approximated described by the following Lorentzian type equation (2)[15].

$$R = \left| 1 - \frac{4 \operatorname{Im}(\beta_0) \operatorname{Im}(\beta_L)}{\{\beta - [\operatorname{Re}(\beta_0) + \operatorname{Re}(\Delta\beta_L)]\}^2 + [\operatorname{Im}(\beta_0) + \operatorname{Im}(\Delta\beta_L)]^2} \right| \quad (2)$$

Equation (2) show that $R_{min} = 0$ will be obtained if $\operatorname{Im}(\beta_0) = \operatorname{Im}(\Delta\beta_L)$, which means the leakage radiation loss is equal to the intrinsic loss. In Figure 3, at the 580nm wavelength $R_{min} = 0$ happens for $TM_1$, $TE_1$ and $TM_2$ modes, so for these mode, $\operatorname{Im}(\beta_0) = \operatorname{Im}(\Delta\beta_L)$. The $[\operatorname{Im}(\beta_0) + \operatorname{Im}(\Delta\beta_L)]$ can be derived out by fitting the ATR curves with equation (2) as shown in Figure 3 (c), (d), (e), which are 0.0042 $K_0$ ($TM_1$), 0.0023 $K_0$ ($TE_1$) and 0.0020 $K_0$ ($TM_2$) respectively. The propagation lengths can be calculated[9] as $0.5*[\operatorname{Im}(\beta_0) + \operatorname{Im}(\Delta\beta_L)]^{-1}$, which are 10.99μm ($TM_1$), 20.67μm ($TE_1$), 23.01μm ($TM_2$). The calculated propagation lengths are a little larger than the experimental results, which can be ascribed to two factors, the first is the scattering from the film roughness induced propagation loss, which has not been taken into account in the theoretical simulations but exists in the real experiments. The second factor is due to the spectrum width of the detected fluorescence. As we know, the LR angles of WMs are dependent on the wavelength. Different wavelength will lead to different LR angles, which results in different diameter of the ring in the Fourier plane image. So if the detected fluorescence contained many wavelengths, the width of the cross section of the ring as shown in Figure 2 (b) (c) (d) will be enlarged, and result in smaller propagation lengths calibrated. While in the theoretical calculation, monochromatic light are used, so the experimental results are a litter smaller than that from theoretical calculation. The measurement precision can be enhanced if the FWHM of the band pass filter was decreased.

In summary, a simple and high efficient method based on the fluorescence and LRM is proposed to characterize the propagation constants of WMs existing in a planar
7

waveguide. There is no need to do the angle scanning or any mechanical movement as used in prism based measurements [16, 17], these propagation constants of WMs can be directly derived from the Fourier plane images. The experimental results and numerical calculation are consistent, which confirm the method. Our work has potential applications in the field of integrated optics because waveguides are the key element in integrated optical circuit.

**Acknowledgment:** DGZ is grateful to X.J. Jiao (University of Utah) for our helpful discussions. This work is supported by the National Key Basic Research Program of China under grant no. 2011CB301802, 2012CB921901, and 2012CB922003. National Natural Science Foundation of China under grant no. 11004182, 61036005, 60977019, and the Fundamental Research Funds for the Central Universities.

FIGURES CAPTIONS:

Figure 1: (a) Fourier plane image of the reflected 532nm laser beam, the three pair of dark arcs are signatures of the excitation of WMs ($TM_1$, $TM_2$, $TE_1$) inside the plane waveguide by linearly polarized light. (b) Fourier plane image of the fluorescence from RhB molecules doped in the waveguide. The four bright rings represent four WMs ($TE_0$,

$TM_1$, $TE_1$ and $TM_2$) excited by emission from RhB molecules at the wavelength of 580nm.

Figure 2: (a) Pixel intensity distribution along the red dash line in Figure 1 (b).The distance between peaks A1 and A2 (d1), B1 and B2 (d2), C1, and C2 (d3) represent the diameter of the corresponding bright ring, from which the wavenumber of WMs can be estimated. (b), (c), (d) the left section of curves in (a) is fitted with Lorentzian curves. The red dots are the experimental data and the blue solid lines are Lorentzian curves used to extract FWHM of the wavenumber distribution in Fourier plane of excited WMs.

Figure 3: (a) (b) Calculated ATR curves at the excitation wavelengths of 532nm and 580nm with TM and TE polarizations. The wavenumbers of excited WMs can be derived from valley positions of the curves. (c) (d) and (e) The ATR curves of $TM_1$, $TE_1$, and $TM_2$ modes ( at 580nm) in the region of resonance are fitted with Lorentzian shape curves. The red dots are the data from the ATR curves in (a) and the blue lines are the Lorentzian curves.

FIGURES

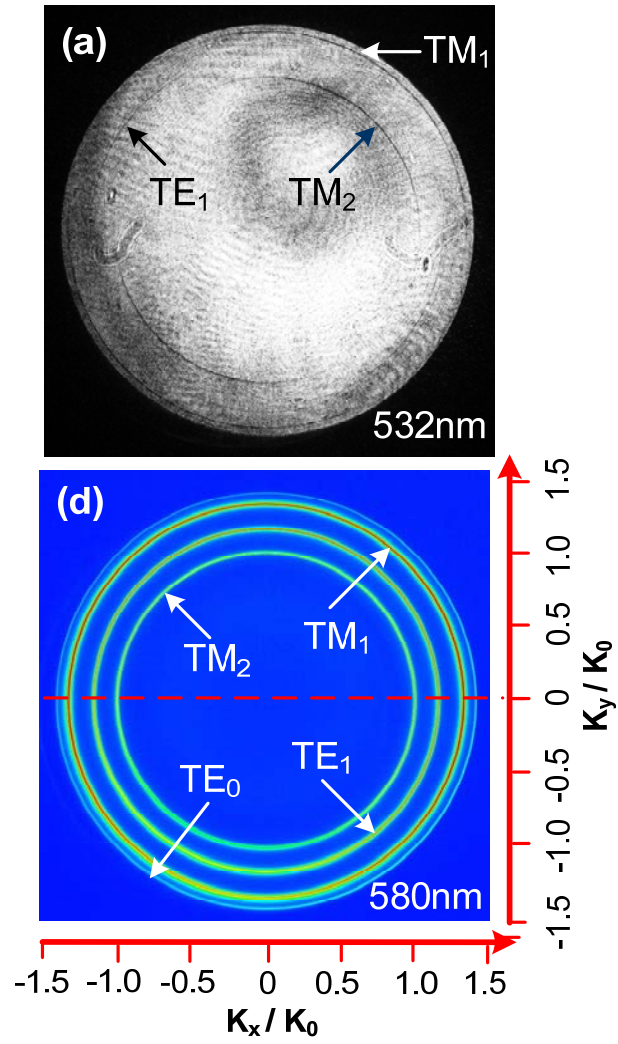

Figure 1

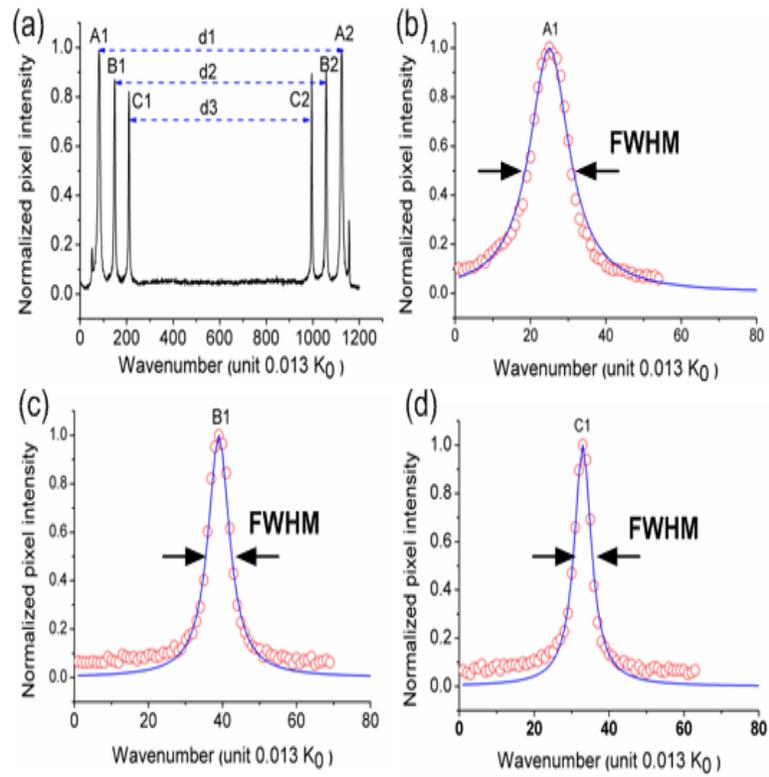

Figure 2

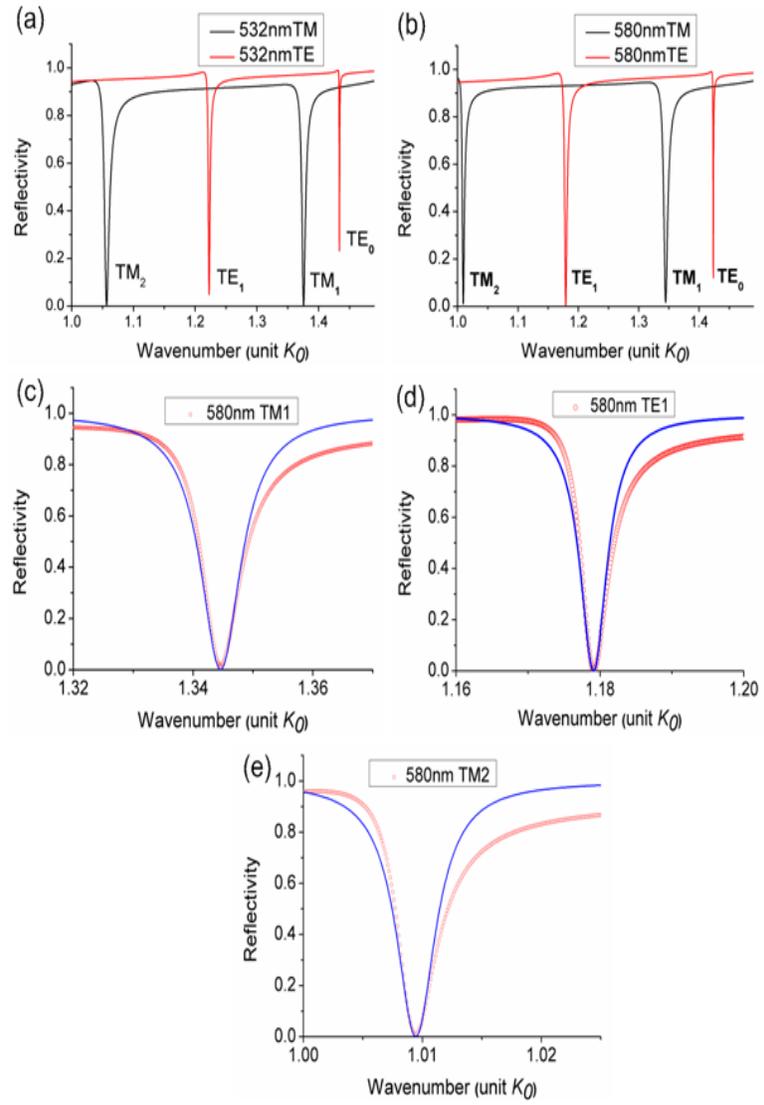

Figure 3